\documentclass[aps, prd, twocolumn,
showpacs, showkeys, preprintnumbers, nofootinbib, superscriptaddress]{revtex4-1}

\usepackage{amssymb,amsbsy,amsmath,amsfonts}
\usepackage{slashed}
\usepackage{bm}
\usepackage{times}

\usepackage{color}

\definecolor{lightyellow}{RGB}{255,250,205}

\usepackage{graphicx}
\graphicspath{{./}{./img/}{./fig/}{./image/}{./figure/}{./picture/}}
\usepackage{epstopdf}

\usepackage{tabularx}
\usepackage{longtable}

\usepackage{soul}
\setulcolor{red} 
\setstcolor{green}
\sethlcolor{lightyellow}

\usepackage{hyperref}
\hypersetup{
  colorlinks=true,        
  linkcolor=blue,         
  citecolor=cyan,         
  urlcolor=red,           
}

\begin{document}

\title{Towards a relativistic formulation of baryon-baryon interactions in chiral perturbation theory
\footnote{\textbf{Foundation item:}
National Natural Science Foundation of China (11375024, 11522539, 11335002, 11621131001); China Postdoctoral Science Foundation (2016M600845, 2017T100008).
}
}

\author{Xiu-Lei Ren}
\affiliation{
State Key Laboratory of Nuclear Physics and Technology,
School of Physics, Peking University, Beijing 100871, China}
\affiliation{Institut f\"{u}r Theoretische Physik II, Ruhr-Universit\"{a}t Bochum, D-44780 Bochum, Germany}

\author{Kai-Wen Li}
\affiliation{
  School of Physics and Nuclear Energy Engineering \& International Research Center for Nuclei and Particles in the Cosmos, Beihang University, Beijing 100191, China}

\author{Li-Sheng Geng}
\email[E-mail: ]{lisheng.geng@buaa.edu.cn}
\affiliation{
  School of Physics and Nuclear Energy Engineering \& International Research Center for Nuclei and Particles in the Cosmos, Beihang University, Beijing 100191, China}
\affiliation{
Beijing Key Laboratory of Advanced Nuclear Materials and Physics, Beihang University, Beijing 100191, China}

\begin{abstract}

In this talk, we report on two recent studies of  relativistic nucleon-nucleon and hyperon-nucleon interactions in covariant chiral perturbation theory, where they are constructed
up to leading order. The relevant unknown low energy constants are fixed by  fitting to
the nucleon-nucleon and hyperon-nucleon scattering data. It is shown that these interactions
can describe the scattering data with a quality similar to their next-to-leading order non-relativistic counterparts. These studies show
that it is technically feasible to construct relativist baryon-baryon interactions, and in addition, after further refinements, these interactions
may provide important inputs to {\it ab initio} relativistic nuclear structure and reaction studies and help improve our understanding of low energy strong interactions.
\end{abstract}

\keywords{nucleon-nucleon interaction, hyperon-nucleon interaction, covariant chiral perturbation theory}

\date{\today}

\maketitle

\section{Introduction}

The nuclear force is  one of the most important inputs to microscopic nuclear structure and reaction studies. It is responsible for holding nucleons together to form nuclei.
As a residual force of the strong interaction, it can in principle be derived from the underlying theory of the strong interaction, Quantum ChromoDynamics (QCD). However, because of the two peculiar properties of QCD, confinement and asymptotic freedom, QCD becomes non-perturbative at the low energy region of nuclear physics interest and, as a result, a direct derivation of realistic nuclear forces from first principles has only become possible in recent years via lattice QCD simulations~\cite{Beane:2010em,Aoki:2012tk}.
In Lattice 2016, the HAL QCD collaboration has reported on their preliminary results of baryon-baryon ($BB$) interactions at the (almost) physical point~\cite{Doi:2017cfx,Sasaki:2017ysy,Ishii:2017xud}.

The original microscopic understanding of the nucleon-nucleon ($NN$) interaction was first proposed by Yukawa, namely, it is mediated by meson exchanges~\cite{Yukawa:1935xg}. Ever since, the idea has been rather popular and successful phenomenologically. Nowadays, there are a variety of  formulations of the nuclear force based on such a picture, such as the high-precision nuclear potentials, Reid93~\cite{Stoks:1994wp}, Argonne $V_{18}$~\cite{Wiringa:1994wb}, (CD-)Bonn~\cite{Machleidt:1989tm,Machleidt:2000ge}. In a similar way, one can derive hyperon-nucleon ($YN$) and hyperon-hyperon ($YY$) interactions as well, such as NSC97a-f~\cite{Rijken:1998yy}and J\"{u}lich 04~\cite{Haidenbauer:2005zh}. These interactions serve as important inputs to (hyper)nuclear structure and reaction studies. Nonetheless, the connection of these phenomenological potentials to  QCD is not very transparent.

The next advancement is the derivation of the nuclear force using chiral perturbation theory (ChPT), which is an effective field theory of low-energy QCD and
 provides a model independent way to study strong-interaction physics~\cite{Weinberg:1978kz}. In 1990s, Weinberg proposed that one can derive the nuclear force from ChPT~\cite{Weinberg:1990rz,Weinberg:1991um}. The so-obtained chiral nuclear forces, based on a consistent power counting scheme, can be systematically improved
 by going to higher orders in terms of external momenta (of the nucleons) and light quark
masses. Three- and four-body interactions can be constructed on the same footing. In recent years, chiral nuclear forces have been
 constructed up to  next-to-next-to-next-to-next-to-leading order (N$^4$LO) and can describe the $NN$ scattering data  with a $\chi^2/\mathrm{datum}\lesssim 1$~\cite{Bedaque:2002mn, Entem:2003ft,Epelbaum:2004fk,Epelbaum:2008ga, Machleidt:2011zz,Epelbaum:2014sza}. In the past decade, the Weinberg approach has been generalized to study antinucleon-nucleon\cite{Kang:2013uia,Dai:2017ont}, $YN$ and $YY$ interactions~\cite{Polinder:2006zh, Haidenbauer:2007ra, Haidenbauer:2013oca, Polinder:2007mp, Haidenbauer:2009qn, Haidenbauer:2015zqb}. Unlike the $NN$ case, the chiral $YN$ and $YY$ interactions have only been formulated up to next-to-leading order (NLO).

The current studies of chiral forces are all based on  non-relativistic (NR) chiral perturbation theory and relativistic effects are either discarded or treated perturbatively. On the other hand, relativistic effects are known to play an important role in understanding the fine structures of atoms/molecules~\cite{Schwerdtfegerbook} and nuclei~\cite{Mengbook}, both of which are conventionally considered as  typical low-energy and non-relativistic systems. Although because of their simplicity, NR approaches are still routinely used in modern studies, the dynamical relativistic effects, such as the appearance of anti-fermions, their spin and the resulting spin-orbit interactions, play a key role in understanding certain properties of
finite nuclei, such as the origin of pseudospin symmetry~\cite{Liang:2014dma}. Furthermore, relativistic effects at the hadronic level have been shown to play an important role as well, by the successful applications of covariant chiral perturbation theory in the one-baryon sector~\cite{Geng:2008mf,Geng:2009ik,Geng:2011wq,Ren:2012aj,Ren:2014vea,Ren:2016aeo} and heavy-light systems~\cite{Geng:2010vw,Geng:2010df,Altenbuchinger:2011qn,Lu:2016kxm}.

Motivated by the successes of relativistic formulations in atomic/molecular, nuclear and hadronic systems and in response to the demand in {\it ab initio} relativistic nuclear structure studies~\cite{Shen:2016bva,Shen:2017vqr}, we proposed  to study the chiral $BB$ interactions in the framework of covariant chiral perturbation theory.  As a first step, we constructed the $NN$ and $YN$ interactions at leading order (LO) and made comparisons with their NR counterparts. The main results will be given in Sec. 3 and Sec. 4, respectively, while more details can be found in Refs.~\cite{Ren:2016jna,Li:2016mln}.

\section{Definition of relativistic baryon-baryon potentials}

A potential is often understood as a quantity used in the non-relativistic
Schrodinger/Lippmann-Schwinger equations. Since our purpose is to construct relativistic
baryon-baryon potentials, it is worthy clarifying the definition of potentials from a
field-theoretical point of view~\cite{Partovi:1969wd,Erkelenz:1974uj}, especially in
the framework of covariant chiral perturbation theory.

Because of the non-perturbative nature of baryon-baryon interactions,
in  relativistic elastic scattering, one has to resort to the
Bethe-Salpeter (BS) equation, which reads, e.g. for nucleon-nucleon scattering,
\begin{eqnarray}
  \mathcal{T}(p',p|W) &=& \mathcal{A}(p',p|W) \nonumber\\
   &+& \int \frac{d^4 k}{(2\pi^4)} \mathcal{A}(p',k|W) G(k|W) T(k,p|W),
\end{eqnarray}
where $p$ ($p'$) is the initial (final) relative four-momentum in the center-of-mass system, and $W=(\sqrt{s}/2,\bm{0})$
is half of the total four-momentum with the total energy $\sqrt{s}=2E_p=2E_{p'}$ and $E_p=\sqrt{\bm{p}^2+m_N^2}$.
$\mathcal{T}$ denotes the invariant amplitude, $\mathcal{A}$ is the interaction kernel consisting of all irreducible
diagrams appearing in covariant chiral perturbation theory (ChPT). $G$ represents the free two-nucleon Green function.
  However, because of both
  formal and practical considerations, one often uses a three-dimensional (3D)
  reduction of the BS equation in practice, such as the Blankenbecler-Sugar
equation~\cite{Blankenbecler:1965gx}, the Thompson equation~\cite{Thompson:1970wt}, the Kadyshevsky equation~\cite{Kadyshevsky:1967rs}, or the Gross equation~\cite{Gross:1969rv}.
In the present work, following Refs.~\cite{Epelbaum:2012ua,Li:2016paq}, we chose to use the Kadyshevsky equation,
\begin{eqnarray}
\label{eq:kadyshevsky}
   T(\bm{p}', \bm{p}) &=&
   V(\bm{p}', \bm{p}) + \int \frac{d^3 k}{(2\pi)^3} \times \nonumber\\
   && \quad V(\bm{p}', \bm{k}) ~ \frac{m_N^2}{2E_k^2} \frac{1}{E_p-E_k + i\epsilon} T(\bm{k}, \bm{p}).
\end{eqnarray}
After integrating out the time component $k_0$ of the BS equation and sandwiching it between the nucleon Dirac spinors, we obtain the relativistic potential, $V$, appearing in the above equation,
\begin{eqnarray}\label{Eq:DefineV}
  V(\bm{p}', \bm{p}) &=& \bar{u}(\bm{p}',s_1) \bar{u}(-\bm{p}',s_2) \times \nonumber\\
  && \mathcal{V}[p'_0=E_{p'}-1/2\sqrt{s},\bm{p}';p_0=E_{p}-1/2\sqrt{s},\bm{p}|W] \times \nonumber\\
  && u(\bm{p}, s_1) u(\bm{p}', s_2),
\end{eqnarray}
with the effective interaction kernel $\mathcal{V}$ perturbatively calculated via
\begin{eqnarray}\label{Eq:mathcalVpert}
  \mathcal{V}^{(2)} &=& \mathcal{A}^{(2)},\nonumber\\
  \mathcal{V}^{(4)} &=& \mathcal{A}^{(4)} + \mathcal{A}^{(2)} (G-g) \mathcal{A}^{(2)},
\end{eqnarray}
and so on, in covariant ChPT.

\section{Relativistic chiral nucleon-nucleon interaction}
In Ref.~\cite{Ren:2016jna}, we proposed a covariant power counting scheme to derive the relativistic chiral nuclear
force defined above. Under this power counting,
we retain the full form of the Dirac spinors, i.e.,
\begin{equation}
  u(\vec{p},s) = N_p
                 \left(\begin{array}{c}
                        1 \\
                        \frac{\vec{\sigma}\cdot\vec{p}}{\epsilon_p}
                 \end{array}\right)\chi_{s}, \quad  N_p=\sqrt{\frac{\epsilon_p}{2M_N}},
\end{equation}
with $\epsilon_p = E_p+M_N$. The chiral dimension of a Feynman diagram is
determined by
\begin{equation}
n_\chi=4L-2N_\pi-N_n+\sum\limits_k k V_k,
\end{equation}
where $L$ is the number of loops, $N_\pi$ the number of internal pion lines, $N_n$ the number of internal nucleon lines, and $V_k$ the vertices of chiral dimension $k$.
In the covariant power counting, the expansion parameters in constructing $V_k$ are the external nucleon three momenta and light quark masses, which are the same as those in the one-baryon sector. It should be mentioned that such power counting schemes are well defined in the $\pi\pi$ and $\pi N$ sectors, but not in the $NN$ sector. Here, we follow the strategies outlined in Refs.~
\cite{Girlanda:2010ya,Djukanovic:2007zz}. In addition, the Dirac spinors are treated as one entity and the small components are retained, different from the NR approach.

According to the above power counting, at leading order one needs to compute the Feynman diagrams shown in Fig.~1.
The relevant Lagrangians are
\begin{equation}
  \mathcal{L}_\mathrm{eff.} = \mathcal{L}_{\pi\pi}^{(2)} + \mathcal{L}_{\pi N}^{(1)} + \mathcal{L}_{NN}^{(0)},
\end{equation}
where the superscript denotes  the chiral dimension. The lowest order $\pi\pi$, $\pi N$ and $NN$ Lagrangians read,
\begin{eqnarray}
  \mathcal{L}_{\pi\pi}^{(2)} &=& \frac{f_\pi^2}{4} \mathrm{Tr}\left[\partial_\mu U \partial^\mu U^\dag + (U+U^\dag) m_\pi^2\right],\\
  \mathcal{L}_{\pi N}^{(1)} &=& \bar{\Psi}\left[i\slashed D - M_{N} + \frac{g_A}{2}\gamma^\mu \gamma_5 u_\mu \right]\Psi,\\
  \mathcal{L}_{NN}^{(0)} &=& \frac{1}{2} \left[C_S (\bar{\Psi}\Psi) (\bar{\Psi}\Psi) + C_A (\bar{\Psi}\gamma_5\Psi) (\bar{\Psi}\gamma_5\Psi)\right.  \nonumber\\
  && + C_V (\bar{\Psi} \gamma_\mu \Psi) (\bar{\Psi} \gamma^\mu\Psi) \nonumber\\
  && + C_{AV} (\bar{\Psi} \gamma_\mu \gamma_5\Psi) (\bar{\Psi}\gamma^\mu\gamma_5\Psi) \nonumber\\
  && + \left. C_T (\bar{\Psi}\sigma_{\mu\nu}\Psi) (\bar{\Psi}\sigma^{\mu\nu}\Psi)\right],
\end{eqnarray}
where the pion decay constant $f_\pi=92.4$ MeV, the axial vector coupling $g_A=1.267$, and $C_{S,A,V,AV,T}$ are low-energy constants (LECs). We note  that the relativistic chiral Lagrangians for nucleon-nucleon interactions have
been constructed up to leading order by a number of groups~
\cite{Polinder:2006zh,Djukanovic:2007zz}.
There are also attempts at the next-to-leading order~\cite{Girlanda:2010ya,Petschauer:2013uua}. Nonetheless, further efforts are still needed to have a systematic power
counting in the relativistic case at NLO and beyond.

As shown in Fig.~1, the LO relativistic chiral nuclear interaction includes four-nucleon contact (CTP) and one-pion-exchange potentials (OPEP),
\begin{equation}
  V_\mathrm{LO}^{NN} = V_\mathrm{CTP}^{NN} + V_\mathrm{OPEP}^{NN},
\end{equation}
\begin{figure}
  \centering
  \includegraphics[width=0.25\textwidth]{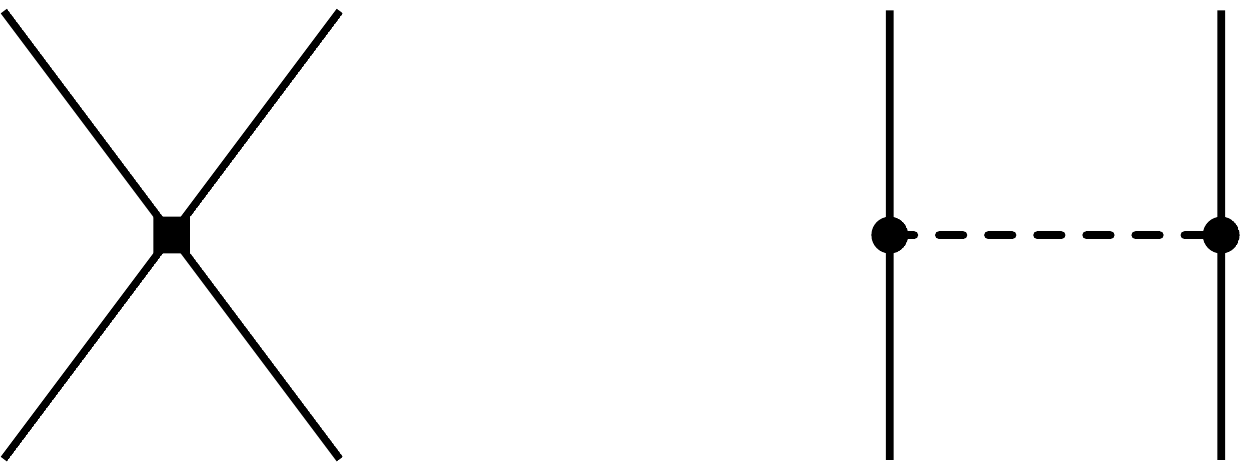}\\
  \caption{Feynman diagrams contributing to the LO relativistic chiral force. Solid lines represent nucleons and the dash line denotes the pion.}
  \label{Fig:FeynLO}
\end{figure}
with
\begin{eqnarray}
   && V_\mathrm{CTP}(\bm{p}',\bm{p}) \nonumber\\
  &=& C_S \left(\bar{u}(\bm{p}', s_1') u(\bm{p}, s_1)\right) \left(\bar{u}(-\bm{p}', s_2') u(-\bm{p}, s_2)\right) \nonumber\\
    &+& C_A \left(\bar{u}(\bm{p}', s_1') \gamma_5 u(\bm{p}, s_1)\right) \left(\bar{u}(-\bm{p}', s_2') \gamma_5 u(-\bm{p}, s_2)\right) \nonumber\\
    &+& C_V \left(\bar{u}(\bm{p}', s_1') \gamma_\mu u(\bm{p}, s_1)\right) \left(\bar{u}(-\bm{p}', s_2') \gamma^\mu u(-\bm{p}, s_2)\right) \nonumber\\
    &+& C_{AV}\left(\bar{u}(\bm{p}', s_1') \gamma_\mu\gamma_5 u(\bm{p}, s_1)\right) \left(\bar{u}(-\bm{p}', s_2') \gamma^\mu\gamma_5 u(-\bm{p}, s_2)\right)  \nonumber\\
    &+& C_{T} \left(\bar{u}(\bm{p}', s_1') \sigma_{\mu\nu} u(\bm{p}, s_1)\right) \left(\bar{u}(-\bm{p}', s_2') \sigma^{\mu\nu} u(-\bm{p}, s_2)\right) ,
\end{eqnarray}
and
\begin{eqnarray}
  V_\mathrm{OPEP}(\bm{p}',\bm{p}) &=&
  -\frac{g_A^2}{4f_\pi^2} \frac{1}
 {(E_{p'}-E_p)^2-(\bm{p}'-\bm{p})^2-m_\pi^2}\nonumber\\
 && \left(\bar{u}(\bm{p}', s_1')\bm{\tau}_1\gamma^\mu\gamma_5q_\mu u(\bm{p}, s_1)\right)\cdot \nonumber\\
 && \left(\bar{u}(-\bm{p}', s_2') \bm{\tau}_2\gamma^\nu\gamma_5q_\nu u(-\bm{p}, s_2)\right),
\end{eqnarray}
where $q$ represents the four momentum transferred $q=(E_{p'}-E_p,\bm{p}'-\bm{p})$ and $\vec{\tau}$ are the isospin Pauli matrices. It should be noted that the retardation effects of OPEP is self-consistently included, consistent with the assumption of the Kadyshevsky equation.

Rewriting the LO potential in terms of the Pauli operators and three momenta, it is easy
to see that $V_\mathrm{LO}^{NN}$ contains all the six allowed spin operators, in contrast with the non-relativistic LO
chiral potential, which only consists of the central, spin-spin and tensor operators.

Next, we perform partial wave decomposition of the chiral potential in the $|LSJ\rangle$ basis and connect them to experimental observables.
First, one calculates the matrix elements of $V_\mathrm{LO}^{NN}$ in the helicity basis, then rotates them to the total
angular momentum space $|JM\rangle$, and finally, transforms them to the $|LSJ\rangle$ representation.
We note that the relativistic contact terms contribute to all the $J \leq 1 $ partial waves and the relativistic corrections to the OPEP are largely suppressed.

  As mentioned before, in the present work, we chose to use the Kadyshevsky equation~\footnote{We checked
    that using the Blankenbecler-Sugar equation to obtain the scattering amplitude does not change our results in any significant way.}, which
  reads in the $LSJ$ basis,
\begin{eqnarray}\label{eq:kadyshevsky}
   T^{SJ}_{L',L}(\bm{p}', \bm{p}) &=& V^{SJ}_{L',L}(\bm{p}', \bm{p}) \nonumber\\
   &+&  \sum\limits_{L''}
   \int_0^{+\infty}\frac{\bm{k}^2 d k}{(2\pi)^3} V^{SJ}_{L',L}(\bm{p}', \bm{k})\frac{M_N^2}{2 E_k^2} \times \nonumber\\
    && \frac{1}{E_p-E_k + i\epsilon} T_{L'',L}^{SJ}(\bm{k}, \bm{p}).
\end{eqnarray}

Furthermore, to avoid ultraviolet divergence, we regularize the potential in Eq.~(\ref{eq:kadyshevsky}) with a form factor. Here, we choose the commonly used separable cutoff function~\cite{Epelbaum:1999dj},
\begin{equation}\label{Eq:formfactor}
  V_\mathrm{LO} \rightarrow V_\mathrm{LO}^\mathrm{Reg.} =
  V_\mathrm{LO}~ \mathrm{exp}\left(\frac{-\bm{p}^{2n}-\bm{p}'^{2n}}{\Lambda^{2n}}\right),
\end{equation}
with $n=2$. One should note that such a form factor is not covariant, but using the same cutoff function as that used in the NR approach allows us to make a direct comparison between the relativistic and NR approaches.\footnote{We realized that
  there exists a covariant but separable cutoff function of the following form,
    \begin{equation}
  V_\mathrm{LO} \rightarrow V_\mathrm{LO}^\mathrm{Reg.} =
  \mathrm{exp}\left[-\left(\frac{p^2-m_N^2}{\Lambda^{2}}\right)\right]~ V_\mathrm{LO}~
  \mathrm{exp}\left[-\left(\frac{{p'}^2-m_N^2}{\Lambda^{2}}\right)\right].
\end{equation}
Preliminary studies show that using such a cutoff function yields slightly better fits compared to what shown in Refs.~\cite{Ren:2016jna}, but the results remain qualitatively similar. More details will be reported in a forthcoming publication.}

In order to determine the five unknown LECs, we need to fit to the $NN$ scattering phase shifts.
We choose the
neutron-proton phase shifts from the Nijmegen93 partial wave analysis~\cite{Stoks:1993tb} with  laboratory kinetic
energy $E_\mathrm{lab.}\leq 100$ MeV.
The momentum cutoff $\Lambda$ is varied from 500 MeV to 1000 MeV.
We found that the
best fitted result is located at $\Lambda=750$ MeV with $\tilde{\chi}^2/\mathrm{d.o.f.} \sim 2.0$, and the corresponding description of phase shifts is presented in
Fig.~2. For the sake of comparison, the results of the LO and NLO non-relativistic chiral force  from
Ref.~\cite{Epelbaum:1999dj} are also shown. Furthermore, the variations from the best fit with the cutoff ranging from 500 MeV to 1000 MeV are shown as the red bands in Fig. 2. The covariant LO results  can better describe the $^1S_0$ and $^3P_0$ phase shifts than the corresponding NR
ones.They are quantitatively similar to the NLO NR ones. It can be seen that the variation of the cutoff does not change qualitatively the overall picture. On the other hand, for the five $J=1$ phase shifts, the relativistic results are almost the same as the non-relativistic ones.

\begin{figure}
\includegraphics[width=0.49\textwidth]{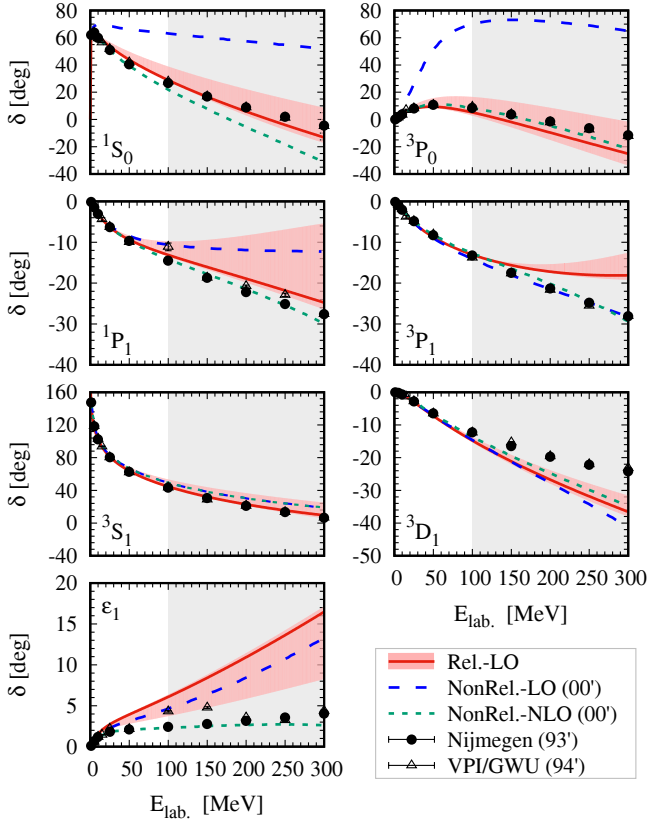}\\
\caption{
Comparison of theoretical and experimental neutron-proton phase shifts for $J\leq1$. The red solid lines represent the results of the LO relativistic potential, while the dashed and dotted lines denote the LO and NLO non-relativistic results~\cite{Epelbaum:1999dj}. The red bands are the relativistic results with the cutoff ranging from $500$ MeV to $1000$ MeV. Solid dots and open triangles are the $np$ phase shifts of Nijmegen~\cite{Stoks:1993tb} and VPI/GWU~\cite{Arndt:1994br}. The gray backgrounds denote the energy regions where the  theoretical results are predictions. The figure is taken from Ref.~\cite{Ren:2016jna}.}
\label{Fig:PSdes}
\end{figure}

\begin{figure}
\includegraphics[width=0.49\textwidth]{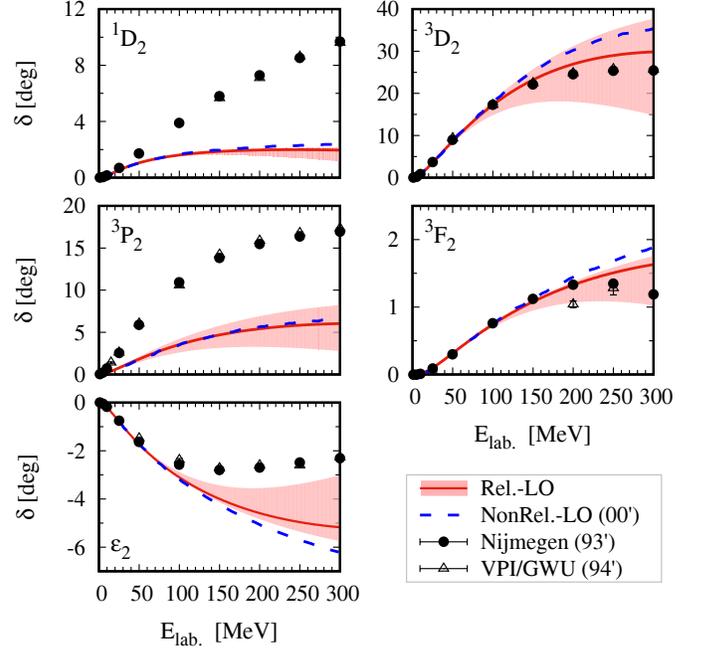}\\
\caption{
Neutron-proton phase shifts for $J=2$. The notations are the same as Fig.~2. The figure is taken from Ref.~\cite{Ren:2016jna}.}
\label{Fig:PSdesJ2}
\end{figure}

In Fig.~3, we show the description of the $J=2$ phase shifts, where only one-pion-exchange diagrams contribute. Following the same strategies, we give the relativistic results with $\Lambda=750$ MeV as central values and the variation bands are obtained by varying the cutoff  from $500$ MeV to $1000$ MeV. For the sake of comparison, the LO non-relativistic results of Ref.~\cite{Epelbaum:1999dj} are also shown. One can see that they are almost the same, because the relativistic corrections of the OPEP are largely suppressed.

\section{Relativistic chiral hyperon-nucleon interaction}
We have extended the covariant power counting scheme to the $YN$ sector and constructed the strangeness $S=-1$ $YN$ interaction up to LO in Ref.~\cite{Li:2016mln}.
The Lagrangians for the contact terms (CT) and the one-pseudoscalar-meson-exchange potential (OPME) read
\begin{align}\label{CT}
  \mathcal{L}_{\textrm{CT}} &=\frac{\tilde C_i^1}{2}~\textrm{tr}\left(\bar B_a \bar B_b (\Gamma_i B)_b (\Gamma_i B)_a\right)
  \notag\\
  &~~+ \frac{\tilde C_i^2}{2}~\textrm{tr}\left(\bar B_a (\Gamma_i B)_a \bar B_b (\Gamma_i B)_b\right)
  \notag\\
  &~~+ \frac{\tilde C_i^3}{2}~\textrm{tr}\left(\bar B_a (\Gamma_i B)_a\right)\textrm{tr}\left( \bar B_b (\Gamma_i B)_b\right),
\end{align}
\begin{align}\label{LMB1}
\mathcal{L}_{MB}^{(1)} &=
  \mathrm{tr}\Bigg( \bar B \big(i\gamma_\mu D^\mu - M_B \big)B
  -\frac{D}{2} \bar B \gamma^\mu\gamma_5\{u_\mu,B\} \notag\\
  &~~  -  \frac{F}{2}\bar{B} \gamma_\mu\gamma_5 [u_\mu,B]\Bigg)\, ,
\end{align}
The potential can be symbolically expressed as
\begin{equation}
  V_\mathrm{LO}^{BB'} = V_\mathrm{CT}^{BB'} + V_\mathrm{OPME}^{BB'}.
\end{equation}
The contact terms are derived assuming SU(3) symmetry~\cite{deSwart:1963pdg}.
The OPME potential reads
\begin{eqnarray}
  V_\mathrm{OPME}^{BB'} &=& -N_{B_1B_3\phi}N_{B_2B_4\phi} \mathcal{I}_{B_1B_2\rightarrow B_3B_4}\nonumber\\
   &&\times\frac{(\bar u_3 \gamma^\mu \gamma_5 q_\mu u_1) (\bar u_4 \gamma^\nu \gamma_5 q_\nu u_2)}
        {q^2-m^2},
\end{eqnarray}
where the SU(3) coefficient $N_{BB'\phi}$ and isospin factor $\mathcal{I}_{B_1B_2\rightarrow B_3B_4}$ can be found in, e.g., Refs.~\cite{deSwart:1963pdg,Polinder:2006zh}. Note that  SU(3) symmetry is broken due to the mass difference of the exchanged mesons. The corresponding Feynman diagrams for the strangeness $S=-1$ $\Lambda N-\Sigma N$ system are shown in Figs.~4-5.

\begin{figure*}
\centering
\includegraphics[width=0.15\textwidth]{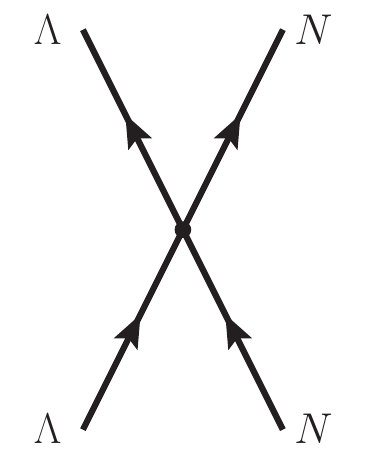}~~
  \includegraphics[width=0.15\textwidth]{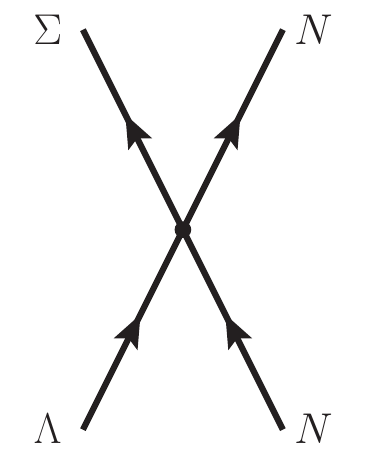}~~
  \includegraphics[width=0.15\textwidth]{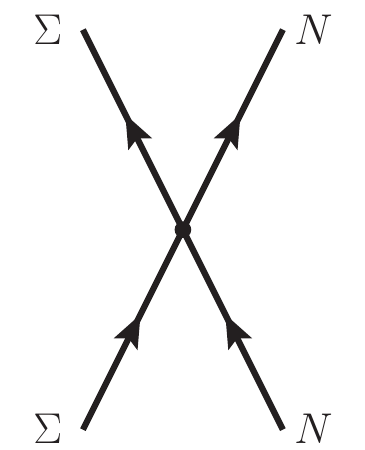}\\
\caption
{Nonderivative four baryon contact terms in the $\Lambda N-\Sigma N$ system.}
\label{Fig:CT3}
\end{figure*}

\begin{figure*}
\centering
  \includegraphics[width=0.15\textwidth]{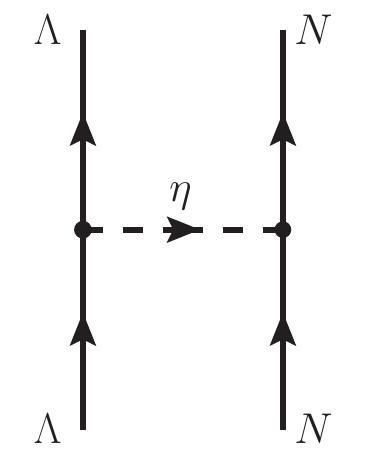}~~
  \includegraphics[width=0.15\textwidth]{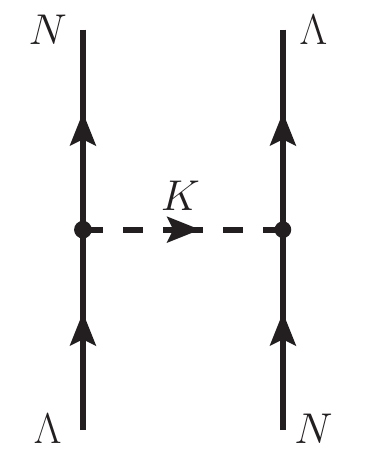}~~
  \includegraphics[width=0.15\textwidth]{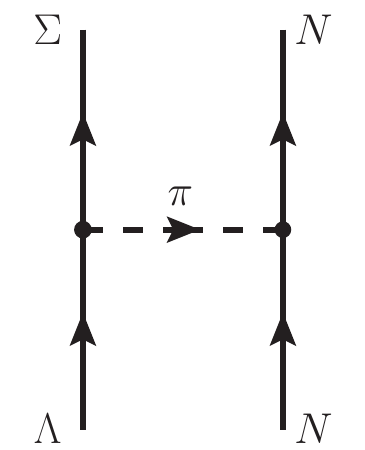}~~
  \includegraphics[width=0.15\textwidth]{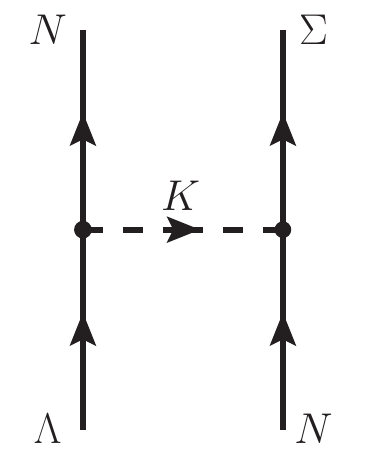}~~
  \includegraphics[width=0.15\textwidth]{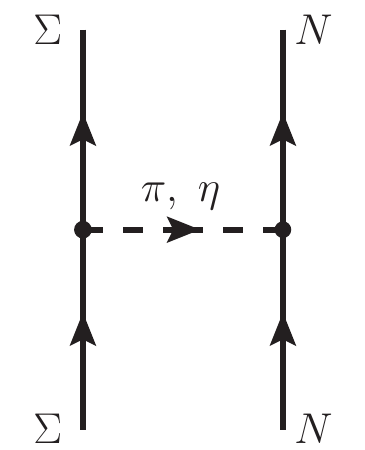}~~
  \includegraphics[width=0.15\textwidth]{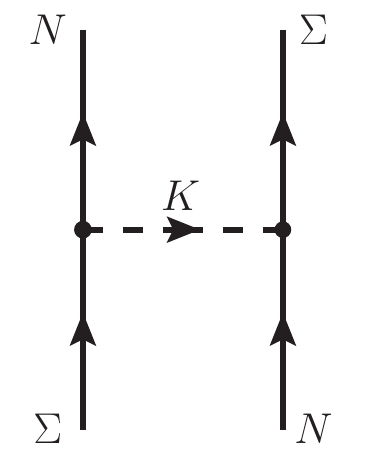}\\
\caption{
One-pseudoscalar-meson exchange diagrams in the $\Lambda N-\Sigma N$ system.}
\label{Fig:OME6}
\end{figure*}


\begin{figure*}
\includegraphics[width=1.0\textwidth]{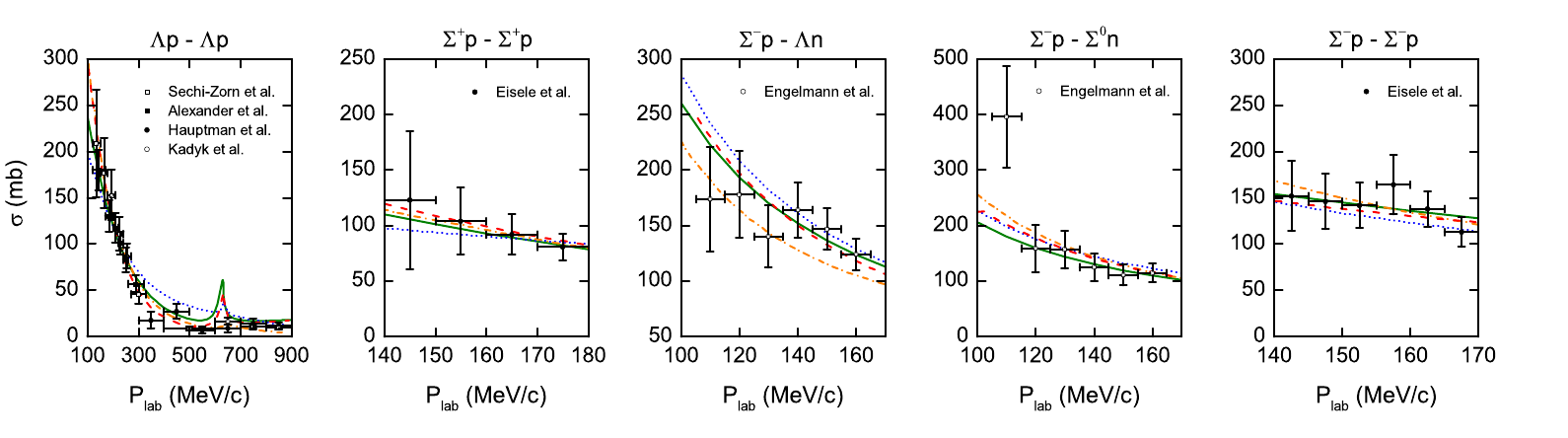}\\
\caption{Cross sections in the leading order relativistic $\chi$EFT approach (green solid lines) and NR(HB) approach (blue dotted lines)as functions of the laboratory momentum at $\Lambda_F=600$ MeV. For reference, the NSC97f \cite{Rijken:1998yy} (red dash lines) and J\"ulich 04~\cite{Haidenbauer:2005zh} (orange dashed-dotted lines) results are also shown. The figure is taken from Ref.~\cite{Li:2016mln}.}
\label{Fig:YNCS}
\end{figure*}

Following the same procedure as in the $NN$ case, one can perform partial wave decompositions of the LO $YN$ potential and obtain $V_\mathrm{LO}^{YN}$ in the $|LSJ\rangle$ basis. After iteration of the partial wave potential in the coupled-channel Kadyshevsky equation,
\begin{align}\label{SEK}
  & T_{\rho\rho'}^{\nu\nu',J}(\mbox{\boldmath $p$}',\mbox{\boldmath $p$};\sqrt{s})
  =
   V_{\rho\rho'}^{\nu\nu',J}(\mbox{\boldmath $p$}',\mbox{\boldmath $p$}) +
  \sum_{\rho'',\nu''}\int_0^\infty \frac{dp''p''^2}{(2\pi)^3} \notag\\
  &\times\frac{M_{B_{1,\nu''}}M_{B_{2,\nu''}}~ V_{\rho\rho''}^{\nu\nu'',J}(\mbox{\boldmath $p$}',\mbox{\boldmath $p$}'')~
   T_{\rho''\rho'}^{\nu''\nu',J}(\mbox{\boldmath $p$}'',\mbox{\boldmath $p$};\sqrt{s})}{E_{1,\nu''}E_{2,\nu''}
  \left(\sqrt{s}-E_{1,\nu''}-E_{2,\nu''}+i\epsilon\right)},
\end{align}
one can obtain the scattering $T$-matrix. In Eq.~(\ref{SEK}), the potential needs also to be regularized to avoid ultraviolet divergence. We chose the Gaussian cutoff function~Eq.~(\ref{Eq:formfactor}) as in the $NN$ case. Furthermore, in order to properly take into account the physical thresholds and the Coulomb force in charged channels, we solve the Kadyshevsky equation in particle basis. The Coulomb effects are treated with the Vincent-Phatak approach.

Among the 15 LECs in the Lagrangian of Eq.~(16),  it  can be easily shown that there are only 12 independent LECs or  equivalently 12 independent partial waves in
the strangeness $S=-1$ sector.~\footnote{The other three LECs  contribute exclusively to the strangeness $S=-2$ system.} To do this, one can write down the following 12 partial wave potentials which are linear functions of the 12 LECs we introduced in Ref.~\cite{Li:2016mln}.
    \begin{align}
    \scriptsize
    & V^{\Lambda\Lambda}_{1S0} = C_{1S0}^{\Lambda\Lambda}\left[1+(R_p^{\Lambda\Lambda})^2(R_{p'}^{\Lambda\Lambda})^2\right] \notag\\
    &\qquad\quad+ \hat C_{1S0}^{BB'}\left[(R_p^{\Lambda\Lambda})^2+(R_{p'}^{\Lambda\Lambda})^2\right], \nonumber\\
    & V^{\Sigma\Sigma}_{1S0} = C_{1S0}^{\Sigma\Sigma}\left[1+(R_p^{\Sigma\Sigma})^2(R_{p'}^{\Sigma\Sigma})^2\right] \notag\\
    &\qquad\quad+ \hat C_{1S0}^{BB'}\left[(R_p^{\Sigma\Sigma})^2+(R_{p'}^{\Sigma\Sigma})^2\right], \nonumber\\
    & V^{\Lambda\Lambda}_{3P1} = -\frac{4}{3}C_{3P1}^{\Lambda\Lambda}R_p^{\Lambda\Lambda}R_{p'}^{\Lambda\Lambda}, \nonumber\\
    & V^{\Sigma\Sigma}_{3P1} = -\frac{4}{3}C_{3P1}^{\Sigma\Sigma}R_p^{\Sigma\Sigma}R_{p'}^{\Sigma\Sigma}, \nonumber\\
    & V^{\Lambda\Lambda}_{3P0} = -2(-C_{1S0}^{\Lambda\Lambda} - \hat C_{1S0}^{\Lambda\Lambda} + 2D_{3S1}^{\Lambda\Lambda} - 2\hat D_{3S1}^{\Lambda\Lambda})R_p^{\Lambda\Lambda}R_{p'}^{\Lambda\Lambda}, \nonumber\\
    & V^{\Sigma\Sigma}_{3P0} = -2(-C_{1S0}^{\Sigma\Sigma} - \hat C_{1S0}^{\Sigma\Sigma} + 2D_{3S1}^{\Sigma\Sigma} - 2\hat D_{3S1}^{\Sigma\Sigma})R_p^{\Sigma\Sigma}R_{p'}^{\Sigma\Sigma}, \nonumber\\
    & V^{\Lambda\Lambda}_{3S1} = C_{3S1}^{\Lambda\Lambda}\left[1+(R_p^{\Lambda\Lambda})^2(R_{p'}^{\Lambda\Lambda})^2\right] \notag\\
    &\qquad\quad+ \hat C_{3S1}^{\Lambda\Lambda}\left[(R_p^{\Lambda\Lambda})^2+(R_{p'}^{\Lambda\Lambda})^2\right], \nonumber\\
    & V^{\Sigma\Sigma}_{3S1} = C_{3S1}^{\Sigma\Sigma}\left[1+(R_p^{\Sigma\Sigma})^2(R_{p'}^{\Sigma\Sigma})^2\right] \notag\\
    &\qquad\quad+ \hat C_{3S1}^{\Sigma\Sigma}\left[(R_p^{\Sigma\Sigma})^2+(R_{p'}^{\Sigma\Sigma})^2\right], \nonumber\\
    & V^{\Lambda\Sigma}_{3S1} = C_{3S1}^{\Lambda\Sigma}\left[1+(R_p^{\Lambda\Sigma})^2(R_{p'}^{\Lambda\Sigma})^2\right] \notag\\
    &\qquad\quad+ \hat C_{3S1}^{\Lambda\Sigma}\left[(R_p^{\Lambda\Sigma})^2+(R_{p'}^{\Lambda\Sigma})^2\right], \nonumber\\
    & V^{\Lambda\Lambda}_{1P1} = -\frac{2}{3}(C_{3S1}^{\Lambda\Lambda} - \hat C_{3S1}^{\Lambda\Lambda})R_p^{\Lambda\Lambda}R_{p'}^{\Lambda\Lambda}, \nonumber\\
    & V^{\Sigma\Sigma}_{1P1} = -\frac{2}{3}(C_{3S1}^{\Sigma\Sigma} - \hat C_{3S1}^{\Sigma\Sigma})R_p^{\Sigma\Sigma}R_{p'}^{\Sigma\Sigma}, \nonumber\\
    & V^{\Lambda\Sigma}_{1P1} = -\frac{2}{3}(C_{3S1}^{\Lambda\Sigma} - \hat C_{3S1}^{\Lambda\Sigma})R_p^{\Lambda\Sigma}R_{p'}^{\Lambda\Sigma},
  \end{align}
   where
   \begin{align}
     & D_{3S1}^{\Lambda\Lambda} = \frac{1}{18}\left(17C_{3S1}^{\Lambda\Lambda}+15C_{3S1}^{\Lambda\Sigma}+C_{3S1}^{\Sigma\Sigma}\right),\notag\\
     & \hat D_{3S1}^{\Lambda\Lambda} = \frac{1}{18}\left(17\hat C_{3S1}^{\Lambda\Lambda}+15\hat C_{3S1}^{\Lambda\Sigma}+\hat C_{3S1}^{\Sigma\Sigma}\right),\notag\\
     & D_{3S1}^{\Sigma\Sigma} = C_{3S1}^{\Lambda\Lambda}+C_{3S1}^{\Lambda\Sigma},\notag\\
     & \hat D_{3S1}^{\Sigma\Sigma} = \hat C_{3S1}^{\Lambda\Lambda}+\hat C_{3S1}^{\Lambda\Sigma}.
   \end{align}
   One can easily check that this set of equations has a unique solution, which means that they are linearly independent.  The remaining potentials, namely the $^3S_1-{}^3D_1$ mixing and $^3D_1$,  can be expressed in terms of $V^{BB'}_{3S1}$ and $V^{BB'}_{1P1}$. The only other choice is to take those LECs in the $^1S_0$, $^3S_1$ and $^3P_0$ partial waves.

To determine the $12$ unknown LECs, we performed a fit to the scarce low energy $YN$ scattering data, which consist of $35$ cross sections and a $\Sigma^-p$ inelastic capture ratio at rest.
We also took into account the $S$-wave scattering lengths of $\Lambda p$ and $\Sigma^+ p$ to further constrain the values of LECs. The cutoff scale $\Lambda$ was varied from $500$ to $850$ MeV. The details of the fit can be found in Ref.~\cite{Li:2016mln}.

The best fitted results are obtained at $\Lambda=600$ MeV, with a $\chi^2=16.1$. The corresponding description of the experimental cross sections are presented in Fig.~6.  For references, the results of two phenomenological potentials, NSC97f and J\"ulich04, and those of the LO NR heavy baryon (HB) chiral force are also shown.
One can see that the relativistic results can reproduce the $YN$ scattering data quite well.

Moreover, differential cross sections are also shown in Fig.~7. One can see that the theoretical predictions agree well with the experimental data within uncertainties, although they are not considered in the fits. $S$- and $P$-wave phase shifts of $\Lambda p$ and $\Sigma^+ p$ reactions are shown in Figs.~8,9. One can see that the $^1S_0$ and $^3P_0$ phase shifts are quite different from those of the LO HB approach, but the $^3P_2$ phase shifts are similar, where only OPME terms contribute.

\begin{figure*}[htpb]
\includegraphics[width=1.0\textwidth]{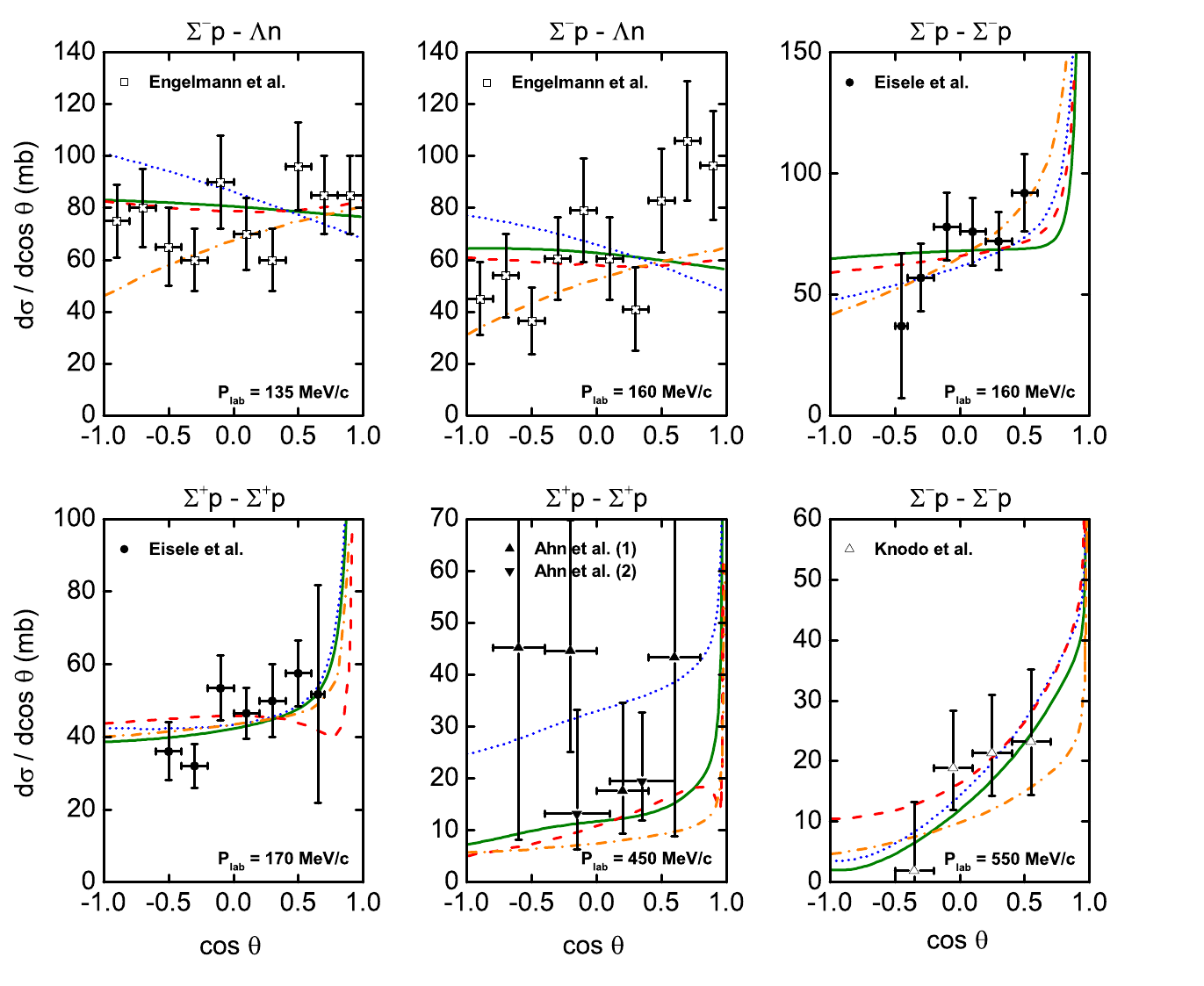}\\
\caption{
Differential cross sections as a function of cos$\theta$ at various laboratory momenta $P_{\textrm{lab}}$, where $\theta$ is the center-of-mass scattering angle. The notations are the same as Fig.~6. The figure is taken from Ref.~\cite{Li:2016mln}.}
\label{dcs}
\end{figure*}

\begin{figure*}[htpb]
\includegraphics[width=1.0\textwidth]{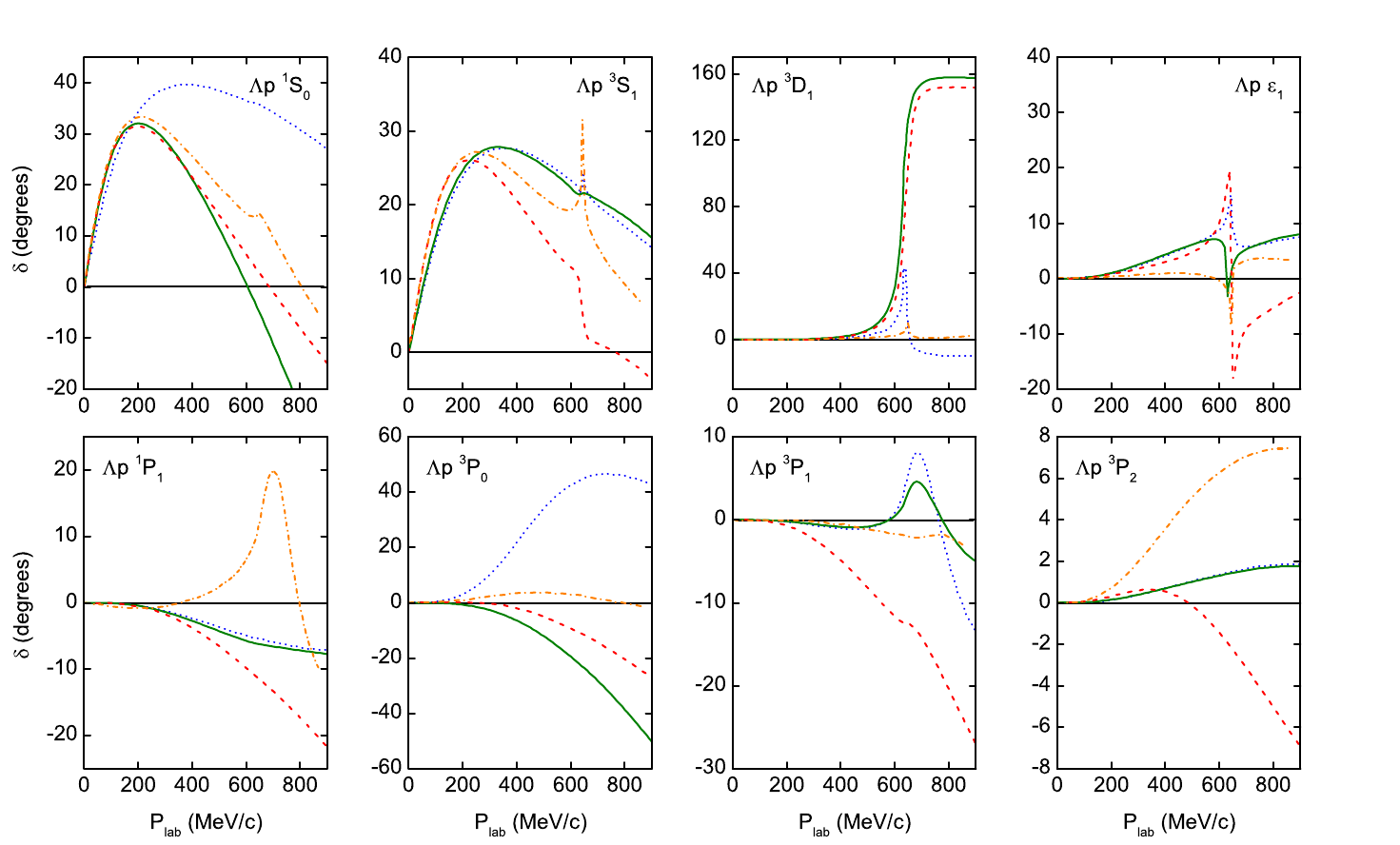}\\
\caption{$\Lambda p$ $S$- and $P$-wave phase shifts in the leading order relativistic ChPT approach (green solid lines) and NR(HB) approach (blue dotted lines) as functions of the laboratory momentum at $\Lambda_F=600$ MeV. For reference, the NSC97f \cite{Rijken:1998yy} (red dash lines) and J\"ulich 04~\cite{Haidenbauer:2005zh} (orange dashed-dotted lines) results are also shown. The figure is taken from Ref.~\cite{Li:2016mln}.}
\label{LaPps}
\end{figure*}

\begin{figure*}[htpb]
\includegraphics[width=1.0\textwidth]{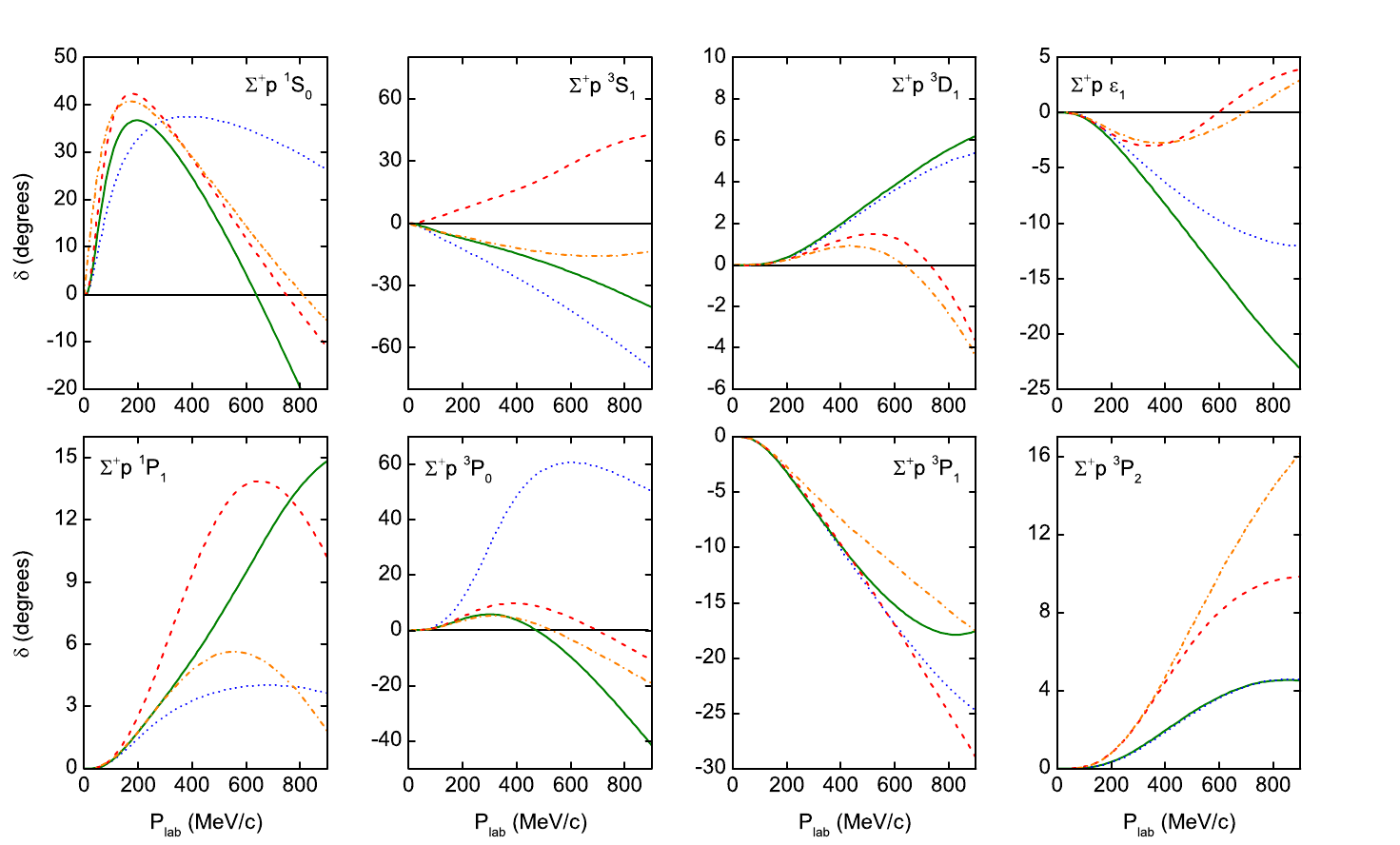}\\
\caption{$\Sigma^+p$ $S$- and $P$-wave phase shifts in various approaches. The notations are the same as in Fig.~8. The figure is taken from Ref.~\cite{Li:2016mln}.}
\label{SiPps}
\end{figure*}

Cutoff dependence of the fitted $\chi^2$ is shown in Fig.~10, in comparison with the LO~\cite{Polinder:2006zh} and NLO~\cite{Haidenbauer:2013oca}  NR approach, and the approach in Refs.~\cite{Epelbaum:2012ua,Li:2016paq} (denoted as the EG approach). The relativistic results are less sensitive to the cutoff variation, compared with the LO NR approach and the EG approach,  and are comparable with the NLO NR approach. Similar to the $NN$ case, the improvement mainly originates from the contact terms.

\begin{figure}
  \centering
  \includegraphics[width=0.5\textwidth]{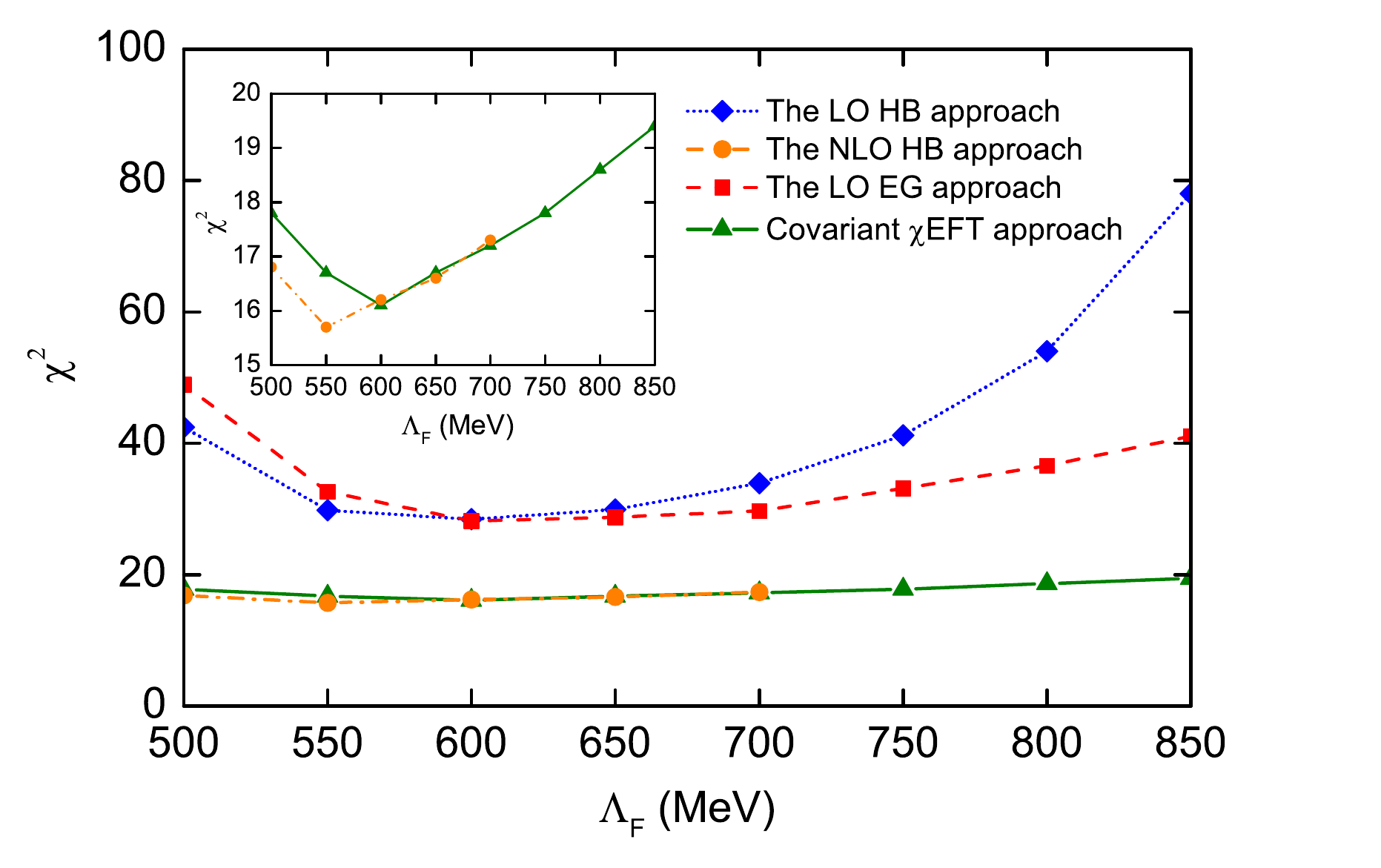}\\
\caption{ $\chi^2$ as a function of the cutoff in the LO (blue dotted line)~\cite{Polinder:2006zh}, NLO (orange dashed-dotted line)~\cite{Haidenbauer:2013oca}  NR(HB) approach, the LO EG approach (red dashed line)~\cite{Li:2016paq} and the LO relativistic $\chi$EFT approach (green solid line).}
  \label{chi2}
\end{figure}

We have tried to describe the $NN$ and $YN$ data simultaneously and found that a simultaneous fit of $NN$ and $YN$ systems is impossible, similar to the NLO NR case~\cite{Haidenbauer:2013oca}. This can be demonstrated in the following way. One can easily see that the LECs in the $NN$ sector fixed by fitting to the Nijmegen partial wave analysis with $E_{\textrm{lab.}}\leq 100$ MeV as described in Sec. 3  are quite different from the ones fixed by fitting to the $YN$ scattering data described above. More specifically, the $^1S_0$ partial waves of $NN (I=1)$ and $\Sigma N (I=3/2)$ share the same SU(3) representation $27$. As a result the contact terms should be the same for these two channels in the SU(3) symmetric limit. However,  we find that the $\Sigma^+p\rightarrow \Sigma^+p$ cross sections are largely overestimated with the LECs determined  from the $NN$ analysis, and even a near-threshold bound state appears. We conclude that SU(3) symmetry breaking must be properly taken into account in order to describe the $NN$ and $YN$ scattering simultaneously.

\section{Summary and perspectives}
 We proposed a new covariant power counting scheme to construct relativistic baryon-baryon ($NN$, $YN$, and $YY$) interactions based on covariant chiral perturbation theory. The $NN$ and $YN$ interactions were formulated up to leading order and  it was shown that they can describe the $NN$ and $YN$ scattering data reasonably well, similar to the next-to-leading order non-relativistic ones. From an effective field theory of point of view, such a feature, namely, being able to describe experiments as relatively low order and with fewer
  low energy constants, is very welcome. Of course, more studies are needed to verify whether it will continue into higher orders.

In the near future, we would like to construct the  relativistic chiral nuclear force up to next-to-next-to-leading order and determine the relevant low-energy constants by fitting to either $NN$ phaseshifts or scattering data directly. We expect to obtain a high precision chiral nuclear force for relativistic nuclear structure and reaction  studies. In
the mean time, we will extend the same framework to study $YN$ and $YY$ interactions. With the latest results from lattice QCD simulations, we can achieve a better determination of the corresponding low energy constants and therefore study the impact of these interactions on various topics of current interests, such as the existence of exotic hadrons and the mass-radius relation of neutron stars.

\begin{acknowledgments}
This work was partly supported by the National Natural
Science Foundation of China (NSFC) under Grants No.
11375024, No. 11522539, No. 11735003 and No. 11775009, by DFG
and NSFC through funds provided to the Sino-German CRC
110 Symmetries and the Emergence of Structure in QCD
(NSFC Grant No. 11621131001, DFG Grant No. TRR110),
the China Postdoctoral Science Foundation under Grants No.
2016M600845, No. 2017T100008, and the Fundamental Research
Funds for the Central Universities.
\end{acknowledgments}

\end{document}